\documentclass[twocolumn,amsmath,amssymb,prl,superscriptaddress]{revtex4-1}
\usepackage{titlesec}
\titleclass{\part}{straight}

\def\vec{\mathbf}

\usepackage[colorlinks,bookmarks=false,citecolor=blue,linkcolor=red,urlcolor=blue]{hyperref}
\usepackage{bm}
\usepackage{epsfig}
\usepackage{color}
\usepackage{graphicx} 
\usepackage{dcolumn}  

\newcommand{\ket}[1]{| #1 \rangle} 
\newcommand{\bra}[1]{\langle #1 |} 

\begin{document}


\title{Magnetic state of pyrochlore Cd$_2$Os$_2$O$_7$ emerging from strong competition of ligand distortions and longer-range crystalline anisotropy}

\author{Nikolay A. Bogdanov}
\affiliation
{Institute for Theoretical Solid State Physics, IFW Dresden, Helmholtzstr. 20, 01069 Dresden, Germany}

\author{R\'emi Maurice}
\affiliation
{Theoretical Chemistry Group, Zernike Institute for Advanced Materials, Rijksuniversiteit Groningen,
Nijenborgh 4, 9747 AG Groningen, The Netherlands}

\author{Ioannis Rousochatzakis}
\affiliation
{Institute for Theoretical Solid State Physics, IFW Dresden, Helmholtzstr. 20, 01069 Dresden, Germany}

\author{Jeroen van den Brink}
\affiliation
{Institute for Theoretical Solid State Physics, IFW Dresden, Helmholtzstr. 20, 01069 Dresden, Germany}
\affiliation{Department of Physics, Technical University Dresden, 01062 Dresden, Germany}

\author{Liviu Hozoi}
\affiliation
{Institute for Theoretical Solid State Physics, IFW Dresden, Helmholtzstr. 20, 01069 Dresden, Germany}

\begin{abstract}
By many-body quantum-chemical calculations, we investigate the role of two structural
effects -- 
local
ligand distortions and the anisotropic Cd-ion coordination -- on
the magnetic state of Cd$_2$Os$_2$O$_7$, a spin $S\!=\!3/2$ pyrochlore.
We find that these effects strongly compete, rendering the magnetic interactions and
ordering crucially depend on 
these geometrical features.
Without trigonal distortions a large easy-plane magnetic anisotropy develops.
Their presence, however, reverses the sign of the zero-field splitting and causes a
large easy-axis anisotropy ($D\!\simeq\!-6.8$ meV), which in conjunction with the
antiferromagnetic exchange interaction ($J\simeq 6.4$ meV) stabilizes an all-in/all-out
magnetic order.
The competition uncovered here is a generic feature of pyrochlore magnets.
\end{abstract}

\maketitle

{\it Introduction.\,}
Oxide compounds of the $5d$ elements are at the heart of intensive experimental and theoretical
investigations in condensed-matter physics and materials science.
The few different structural varieties displaying Ir ions in octahedral coordination and
tetravalent $5d^5$ valence states, in particular, have set the playground for new insights
into the interplay between strong spin-orbit interactions and electron correlation effects
\cite{214Ir_kim_2009,227Ir_pesin_2010,kitaev_213Ir_chaloupka10}.
Equally intriguing but less investigated are pentavalent Os $5d^3$ oxides such as the
pyrochlore Cd$_2$Os$_2$O$_7$ \cite{Os227_mandrus_2001,Os227_matsuda_2011}
and the perovskite NaOsO$_3$ \cite{NaOsO3_shi_2009,NaOsO3_calder_2012}.
Just as for the iridates, major open questions with respect to the osmates are the mechanism of the 
metal to insulator transition (MIT) and the nature of the low-temperature antiferromagnetic
(AF) configuration. 
While initially a Slater type picture has been proposed for the MITs in both Cd$_2$Os$_2$O$_7$
\cite{Os227_mandrus_2001} and NaOsO$_3$ \cite{NaOsO3_shi_2009,NaOsO3_calder_2012}, alternative
scenarios such as a Lifshitz transition in the presence
of sizeable electron-electron interactions have been suggested as well very recently
\cite{Os227_shinaoka_2012,Os227_yamaura_2012}.

So far theoretical investigations of the electronic structure of the Os $5d^3$ oxides have been
carried out only as mean-field ground-state (GS) calculations within the local-density approximation
(LDA) \cite{Os227_singh_2002} or the LDA+$U$ model \cite{NaOsO3_shi_2009,Os227_shinaoka_2012}.
Here, we report results of many-body quantum-chemical calculations for Cd$_2$Os$_2$O$_7$.
We describe the local Os $d^3$ multiplet structure, the precise mechanism of 2nd-order spin-orbit
coupling (SOC) and zero-field splitting (ZFS),
and determine the parameters of the effective spin Hamiltonian, i.e.,
the single-ion anisotropy (SIA), nearest-neighbor (NN) Heisenberg exchange as well as
the Dzyaloshinskii-Moriya (DM) interactions. 
The results indicate that basic electronic-structure parameters such as the SIA
crucially depend on geometrical details concerning both the O and adjacent Cd ions.
In particular, the anisotropic Cd-ion coordination lifts the degeneracy of the Os $t_{2g}$ levels
even in the absence of O-ion trigonal distortions and yields large ZFSs and
easy-plane magnetic anisotropy.
A configuration with trigonally distorted O octahedra is however energetically more favorable,
changes the sign of the ZFS, and thus gives rise to easy-axis anisotropy at the Os sites,
which in conjunction with the AF NN exchange stabilizes the so-called all-in/all-out spin order.
The competition between the two structural effects is a generic feature in 227 pyrochlores
and opens new perspectives on the basic magnetism in these materials.

\begin{figure}[lb!]
\includegraphics[width=0.95\columnwidth]{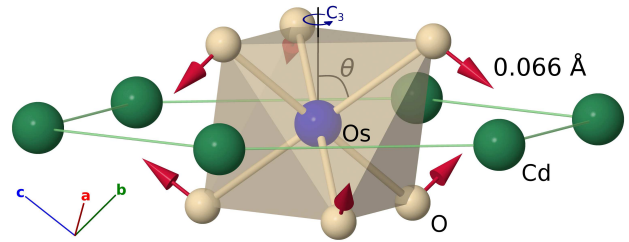}
\caption{Sketch of (i) the compressive trigonal distortion in Cd$_2$Os$_2$O$_7$
(red arrows) involving the O ligands (in beige) 
around each Os site (blue)
and (ii) the hexagonal Cd-ion (green) coordination.
The clusters used in the calculation of the ZFSs also include the six NN octahedra, see text.}
\label{distortions}
\end{figure}

{\it Os $d^3$ electron configuration, single-ion physics.\,}
To investigate in detail the Os $d$-level electronic structure, multiconfiguration self-consistent-field
(MCSCF) and multireference configuration-interaction (MRCI) calculations \cite{bookQC_2000,*fulde_new_book}
were first performed on embedded clusters made of one reference OsO$_6$ octahedron, six adjacent Cd
sites, and six NN OsO$_6$ octahedra.
The farther solid-state environment was modeled as an one-electron effective potential which in
a ionic picture reproduces the Madelung field in the cluster region.
All calculations were performed with the {\sc molpro} quantum-chemical software \cite{molpro_brief}.
We employed energy-consistent relativistic pseudopotentials for Os \cite{ECP_Stoll_5d} and Cd
\cite{ECP_Dolg_4d} and Gaussian-type valence basis functions from the {\sc molpro} library
(see Supplementary Material \cite{Os227_supmat}).

From simple considerations based on crystal-field (CF) theory, the orbital GS for a $d^3$ ion
in octahedral coordination is singlet. 
Thus, neglecting SOC, the lowest electron configuration should be $^4\!A_{2}$($t_{2g}^3$).
This is indeed found in the quantum-chemical calculations if SOC is not included. 
The components of the spin-quartet $A_{2}$ GS can interact, however, via SOC with
higher lying $T_{2}$ terms and for noncubic axial systems split into two Kramers doublets, 
$m_S\!=\!\pm 1/2$ and $m_S\!=\!\pm 3/2$, respectively \cite{ZFS_boca_2004,ZFS_d3_macfarlane67,*ZFS_d3_macfarlane63}.
The $T_{2}$ states are themselves split in noncubic CFs. 
In the recent work of Matsuda {\it et al.}~\cite{Os227_matsuda_2011}, for instance, the interaction
of the $^4\!A_{2}$ and lowest $^4T_{2}$ states was analyzed.
The {\it ab initio} quantum-chemical calculations yield, nevertheless, a rather large $^4\!A_{2}$--$^4T_{2}$
excitation energy, 5.4 eV by MCSCF calculations including in the active orbital space all $5d$
functions at the central Os site and the $t_{2g}$ orbitals of the six Os NNs, which
is not surprising in fact for extended $5d$ orbitals that feel very effectively the O $2p$ charge
distribution. 
Additionally, the {\it ab initio} investigation actually shows that 
the excited $^2\!E$ and $^2T$ $t_{2g}^3$ doublets occur at much lower energies, 1 to 2.5 eV
(see Table~\ref{dd_actual}), and the ZFS of the
$^4\!A_{2}$ levels is mainly related to the interaction with those states. 
To make the MRCI calculations feasible
\footnote{the size of the MRCI expansion is that way reduced by a few orders of magnitude}
and the analysis of the results limpid,
we further replaced the six Os$^{5+}$ $d^3$ NNs by closed-shell Ta$^{5+}$ $d^0$ species.
We checked, for example, that the $5d$ charge distribution and the $t_{2g}$--$e_g$ splitting
at the central Os site are not affected by this substitution.
In the MRCI computations, we included on top of the MCSCF wave functions all single and double 
excitations from the Os $5d$ and O $2p$ orbitals at the central octahedron.
The reference multiconfiguration active space is given by five $5d$ orbitals at the central Os site and three electrons
and the results are listed in Table~\ref{dd_actual}.

\begin{table}[!t]
\caption{
MRCI and MRCI+SOC relative energies (eV) for the Os$^{5+}$ $5d^3$ multiplet structure in Cd$_2$Os$_2$O$_7$.
Since cubic symmetry is lifted, the $T$ states are split even without SOC. 
Each MRCI+SOC value stands for a spin-orbit doublet; for the $^4T$ states only the lowest and highest components are shown.}
\label{dd_actual}
\begin{ruledtabular}
\begin{tabular}{lll}
$5d^3$ splittings             &MRCI                 &MRCI+SOC ($\times 2$)  \\
\hline
\\
$^4\!A_{2}$ ($t_{2g}^3$)      &0.00                 &0.00; $13.5\cdot 10^{-3}$ \\
$^2E$       ($t_{2g}^3$)      &1.51; 1.51           &1.40; 1.53    \\
$^2T_{1}$   ($t_{2g}^3$)      &1.61; 1.62; 1.62     &1.63; 1.66; 1.76 \\
$^2T_{2}$   ($t_{2g}^3$)      &1.46; 2.49; 2.49     &2.63; 2.76; 2.87 \\
$^4T_{2}$   ($t_{2g}^2e_g^1$) &5.08; 5.20; 5.20     &5.14; ...\,; 5.45   \\
$^4T_{1}$   ($t_{2g}^2e_g^1$) &5.89; 6.01; 6.01     &6.02; ...\,; 6.33    \\
$^4T_{1}$   ($t_{2g}^1e_g^2$) &10.29; 10.63; 10.63  &10.41; ...\,; 11.00 \\
\end{tabular}
\end{ruledtabular}
\end{table}

To extract the ZFS, we compute the spin-orbit interaction matrix for the lowest four 
quartet and three doublet spin states, see Table~\ref{dd_actual}. 
The MRCI+SOC data of Table~\ref{dd_actual} show that the ZFS of the $^4\!A_{2}$ GS is 
rather large, $13.5$ meV.
This gives a first estimate of the splitting parameter $|D|\!\approx\!6.75$ meV, one
to two orders of magnitude larger than for $3d$ ions in standard coordination
\cite{ZFS_boca_2004,ZFS_d3_macfarlane67,ZFS_d3_macfarlane63}.
However, for explicitly deriving the full SIA tensor $\bar{\bar{\vec{D}}}$, we further employ the effective 
Hamiltonian methodology described in Ref.~\cite{Heff_Maurice_2009}.
Here, SOCs and the mixing of the $^4\!A_{2}$ components with higher lying CF excited
states are treated as small perturbations and the spin-orbit wave functions related
to the high-spin $t_{2g}^3$ configuration are projected onto the space spanned by the
$^4\!A_{2}$\, $\ket{S, M_S}$ states.
Using the orthonormalized projections $\tilde{\Psi}_k$ of the low-lying $A_{2}$ quartet wave functions
and the corresponding eigenvalues $E_k$ we then construct the effective Hamiltonian
$\mathcal{\hat{H}}_{\mathrm{eff}}\!=\! \sum_{k} E_k \ket{\tilde{\Psi}_k} \bra{ \tilde{\Psi}_k}$
\cite{Heff_Maurice_2009}.
The orthonormalization is done using the formalism introduced by des Cloizeaux \cite{Ornorm_desCloizeaux_1960}, 
i.e., $\ket{\tilde{\Psi}_k}\!=\!\vec{U}_{\mathrm{ov}}^{-\frac{1}{2}} \ket{\Psi_{k}}$,
where $\vec{U}_{\mathrm{ov}}$ is the overlap matrix and $\ket{\Psi_{k}}$ are the projections of the wave functions onto the model space. 
A one-to-one correspondence can be now drawn between the matrix elements of $\mathcal{\hat{H}}_{\mathrm{eff}}$ and
the model Hamiltonian for the anisotropic single site problem, 
$\mathcal{\hat{H}}_{\mathrm{mod}} = \vec{S}\cdot\bar{\bar{\vec{D}}}\cdot\vec{S}$, 
to extract the ZFS tensor 
(for details, see Supplementary Material \cite{Os227_supmat}).
Diagonalizing the $\bar{\bar{\vec{D}}}$ tensor we finally obtain on the basis of the {\it ab initio}
quantum-chemical data an axial parameter $D\!=\!-6.77$ meV, a rhombic parameter $E\!=\!0.0$,
and an univocally defined easy axis of magnetization 
along the $[111]$ direction, 
perpendicular to the plane of Cd NNs. 

To gain deeper insight into the origin of the strong axial anisotropy, we performed additional calculations
for an ideal pyrochlore structure with no trigonal distortions of the O octahedra.
It turns out that the 227 pyrochlore lattice is fully defined by
just three parameters: the space group, the cubic lattice constant $a$, and the 
coordinate $x$ of the O at the $48f$ site, denoted O(1) in Ref.~\cite{Os227_mandrus_2001}. 
For $x\!=\!x_0\!=\!5/16$, the oxygen cage around each Os site forms an undistorted, regular
octahedron \cite{A2B2O7_rao_1983}.
In Cd$_2$Os$_2$O$_7$, however, $x$ is slightly larger than $x_0$ \cite{Os227_mandrus_2001},
which translates into a compressive trigonal distortion of the OsO$_6$ octahedra.
Interestingly, for undistorted O octahedra (i.e., $x\!=\!x_0$) we still found $E\!=\!0$ and the
same orientation of the magnetic $z$ axis.
The axial parameter, however, changes its sign to $D\!=\!2.50$ meV.
A positive $D$ indicates that the $z$ axis is now a hard axis.
The basic mechanism responsible for this change of sign is apparent when comparing the
Os $5d^3$ multiplet structure for the actual lattice (Table~\ref{dd_actual}) with that for a
idealized crystal with no trigonal distortions (Table~\ref{dd_5.16}).
In particular, the order of the split components of the $^2T_{1}$ states,
i.e., one singly-degenerate and one doubly-degenerate term, changes in Table~\ref{dd_5.16} as
compared to Table~\ref{dd_actual}, which modifies the interaction of the $A_{2}$ terms with the excited
states.

\begin{table}[t]
\caption{MRCI and MRCI+SOC relative energies (eV)
for the Os$^{5+}$ $5d^3$ multiplet structure in a hypothetical crystal with no trigonal distortions. 
Each MRCI+SOC value stands for a spin-orbit doublet; 
for the $^4T$ states only the lowest and highest components are shown.}
\label{dd_5.16}
\begin{ruledtabular}
\begin{tabular}{lll}
$5d^3$ splittings             &MRCI                  &MRCI+SOC ($\times 2$)  \\
\hline
\\
$^4\!A_{2}$ ($t_{2g}^3$)      &0.00                  &0.00; $5.00\cdot 10^{-3}$ \\
$^2E$       ($t_{2g}^3$)      &1.53; 1.53            &1.43; 1.50    \\
$^2T_{1}$   ($t_{2g}^3$)      &1.61; 1.61; 1.63      &1.64; 1.64; 1.75 \\
$^2T_{2}$   ($t_{2g}^3$)      &2.42; 2.45; 2.45      &2.59; 2.75; 2.81 \\
$^4T_{2}$   ($t_{2g}^2e_g^1$) &5.34; 5.41; 5.41      &5.36; ...\,; 5.65    \\
$^4T_{1}$   ($t_{2g}^2e_g^1$) &6.13; 6.20; 6.20      &6.23; ...\,; 6.52    \\
$^4T_{1}$   ($t_{2g}^1e_g^2$) &10.85; 10.98; 10.98   &10.93; ...\,; 11.38  \\
\end{tabular}
\end{ruledtabular}
\end{table}

\begin{figure}[b!]
\includegraphics[width=0.90\columnwidth]{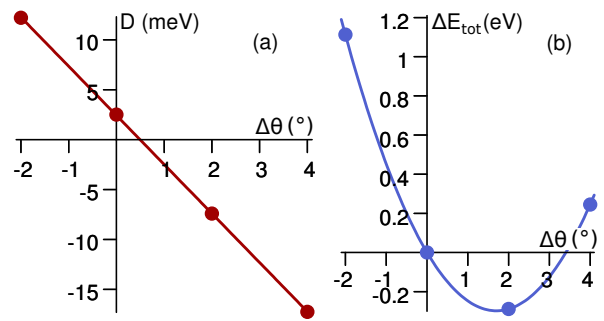}
\caption{(a) Single-ion anisotropy parameter $D$ 
and (b) total energy of the cluster $\Delta E_{\mathrm{tot}}$, as functions of
the trigonal distortion angle $\Delta\theta$, see text.
MRCI+SOC results for clusters as that depicted in Fig.~\ref{distortions}.
No trigonal distortions are applied within the embedding,
i.e., $x\!=\!x_0\!=\!5/16$ for the surrounding lattice.}
\label{Dtheta}
\end{figure}

We find in the MRCI calculations a linear relation between $D$ and the distortion angle
$\Delta\theta$, see Fig.~\ref{Dtheta}.
The effect of a trigonal field alone on the energy levels and on the ZFS of $d^3$ ions 
has been previously investigated in the framework of a simplified CF model and
perturbation theory \cite{ZFS_d3_macfarlane67,ZFS_d3_macfarlane63,ZFS_d3_li09}.
A linear relation is also established in alum salts \cite{ZFS_d3_li09}.
In contrast to the model assumed in Ref.~\cite{ZFS_d3_li09}, however, we find that the
$D(\Delta\theta)$ line does not go through origin.
This is due to the noncubic field generated by the Cd NNs, which form a highly anisotropic planar 
structure arround each Os site.
Similar competing stuctural effects have been evidenced in Ni$^{\mathrm{II}}$--Y$^{\mathrm{III}}$ molecular 
complexes \cite{remi_NiY_2011}.
We also plot in Fig.~\ref{Dtheta} the total-energy landscape as function of the distortion angle
$\Delta\theta$.
It is seen that at the MRCI level the minimum of the parabola corresponds to a trigonal
compressive distortion $\Delta\theta\!=\!1.7^{\circ}$ for fixed Os-O bond length, close to the
experimental value of $1.9^{\circ}$ \cite{Os227_mandrus_2001}.

\begin{figure}[b!]
\includegraphics[width=0.9\columnwidth]{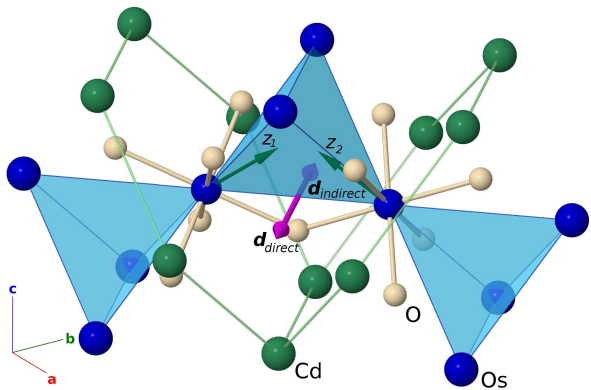}
\caption{Network of corner-shared Os$_4$ tetrahedra (blue color) in Cd$_2$Os$_2$O$_7$.
O ligands and NN Cd ions around two adjacent Os sites are shown in beige and green.
The two possible orientations of the DM vector \cite{pyrochlores_elhajal_05}
for two adjacent Os ions is also shown, along with their respective easy axes.
}
\label{DM}
\end{figure}

{\it Superexchange interactions.\,}
To investigate the NN magnetic couplings, we designed ten-octahedra embedded clusters.
The analysis of the intersite interactions is here carried out for only two magnetically
active Os $d^3$ ions (see Fig. \ref{DM}), while for simplicity 
the eight Os$^{5+}$ NNs were modeled as closed-shell Ta$^{5+}$ $d^0$ ions.
Multiconfiguration wave functions were first generated by state-averaged MCSCF optimizations for 
the lowest singlet, triplet, quintet, and septet states in the two-site problem. 
Those configuration state functions give rise in the spin-orbit calculations to sixteen spin-orbit states. 
For each spin multiplicity the MCSCF active space is defined by the set of six Os $t_{2g}$ orbitals 
accommodating a total number of six electrons. 
We then further accounted for single and double excitations from the Os $t_{2g}$ and bridging-O 
$2p$ orbitals on top of the MCSCF reference wave functions. 
Such MRCI calculations yield magnetic coupling constants in very good agreement with experimental data 
in the $5d^5$ iridate Sr$_2$IrO$_4$ \cite{214Ir_vmk_2012}. 
Similar strategies of explicitly dealing only with selected groups of localized ligand orbitals
were adopted in earlier studies on $3d$ compounds \cite{QC_J_fink94,*J_ligand_calzado03,*NOCI_J_hozoi03}.

From the calculations without SOCs, we see that the energy splittings between the different spin states follow the 
sequence $J$, 3$J$, 6$J$ and can be fitted to an AF isotropic exchange model
$\mathcal{\hat{H}}_{\mathrm{AF}}\!=\!J\,{\bf S}_1\!\cdot\!{\bf S}_2$,
with a mean squared error of $0.6$ meV. 
Adding a biquadratic term $K(\vec{S}_{1}\cdot\vec{S}_{2})^2$ improves the accuracy of the fit
to $0.01$ meV and yields a first MRCI estimate of $J\!=\!6.42$ meV.
Thus $J$ is of the same order of magnitude as the SIA while $K\!=\!0.07$ meV.

{\it Antisymmetric exchange.\,}
Anisotropic, non-Heisenberg terms can be further obtained from spin-orbit calculations
within the manifold defined by the lowest singlet, triplet, quintet, and septet spin states for two Os $d^3$ sites.  
As expected, including SOCs does not bring significant corrections to the 
effective superexchange: $J$ is $6.43$ meV by MRCI+SOC calculations. 
The spin-orbit interactions, however, lift the degeneracy of the initial spin 
multiplets and brings in antisymmetric contributions. 
To describe the latter, we adopt a two-site model Hamiltonian $\mathcal{\hat{H}}_{\text{mod}}'$
containing, in addition to the Heisenberg and biquadratic terms, a DM antisymmetric
exchange contribution $\vec{d}\!\cdot\!\vec{S}_{1}\!\times\!\vec{S}_{2}$,  
where $\vec{d}$ is a pseudovector with components $d_{x}$, $d_{y}$, $d_{z}$.
This model is then compared with the {\it ab initio} $16\!\times\!16$ effective matrix
$\mathcal{\hat{H}}_{\mathrm{eff}}'$ derived from the MRCI+SOC calculations \cite{Os227_supmat}.
We found perfect one-to-one correspondence between the two sets of matrix elements and could extract
the numerical parameters $J\!=\!6.43$, $K\!=\!0.07$, and $\vec{d}\!=\!(1.17, -1.17, 0)$ meV.
According to these results, the DM vector is oriented along the $\langle 110\rangle$
axes (see Fig.~\ref{DM} and Ref.~\cite{pyrochlores_elhajal_05}) and its norm is
$|\vec{d}|=1.65$ meV.
The sense of the DM vector for a given Os-Os link, however, cannot be determined from our {\it ab intio} data.

{\it Discussion.\,}
Whereas the absolute value of the MRCI DM parameter is similar to estimates derived for certain 
values of the on-site Coulomb repulsion $U$ by LDA+$U$ calculations \cite{Os227_shinaoka_2012}, 
the superexchange $J$ and the SIA $D$ turn out to be two ($J$) to seven ($D$) times 
smaller in our quantum-chemical study.
It is known that quantum-chemical calculations based on either configuration-interaction
techniques or 2nd-order perturbation theory are able to reproduce the experimentally derived
$J$ coupling constants with an accuracy better than $\pm 20\%$ in insulating $3d$-metal oxides, see, e.g.,
Refs.~\cite{CuO2_J_munoz_02,*NiO_J_degraaf_97,*V2O5_J_hozoi_03} and \cite{QC_J_fink94,*J_ligand_calzado03,*NOCI_J_hozoi03}.
For more extended $5d$ electronic states in $5d$-metal oxides, the Hubbard $U$ is neither small nor large
compared to the bandwidth, as corroborated from the MITs \cite{Os227_mandrus_2001,NaOsO3_calder_2012}
and from constrained LDA calculations \cite{NaOsO3_shi_2009,Os227_shinaoka_2012,227Ir_savrasov_2011}.
Electron itineracy may therefore pose additional technical problems and limit the applicability of a
finite cluster approach.
It has been shown, however, that
accurate $J$'s \cite{214Ir_vmk_2012} and $d$-$d$ excitation energies
\cite{Ir3116_rixs_liu_2012,*Ir213_rixs_gretarsson_2012,*Ir227_rixs_hozoi_2012} can be indeed
computed for $5d$ oxides such as the iridates.
Those findings in \cite{214Ir_vmk_2012} and \cite{Ir3116_rixs_liu_2012,*Ir213_rixs_gretarsson_2012,*Ir227_rixs_hozoi_2012}
indicate that approaching the essential physics in the $5d$ oxides from a
more localized perspective and within a finite-fragment framework certainly is a good starting point.
Additional potentially problematic points concern the size of the clusters, as discussed, e.g., in
Ref.~\cite{CuO2_dd_hozoi_2011}.
The clusters we employ here are, however, large enough to ensure 
an accurate description of both the charge distribution of NN octahedra around the
``active'' $d$ sites and the tails at those neighboring sites of Wannier-like orbitals in
the active region.
As concerns the SIA at $d$-metal sites in solids, extensive quantum-chemical investigations are missing.
The present investigation therefore constitutes a step toward filling this gap.

The pure AF Heisenberg model on the pyrochlore lattice has an extensive number of classical GSs 
which prevent any N\'eel type ordering down to zero temperature 
\cite{Reimers_1992,*Reimers_1991,Moessner_prl1998,*Moessner_prb1998}.  
The easy-axis anisotropy lifts this macroscopic degeneracy and selects the so-called all-in/all-out (or it's time reversed) state, whereby 
all four spins sharing a given tetrahedron point toward or away from its center.   
This GS is selected irrespective of the relative strength of the anisotropy and AF exchange energy
\cite{227_moessner_98}.  
Now, since the macroscopic GS degeneracy is already lifted by the SIA, including 
the smaller DM interactions is not expected to drastically modify the physics of the system. 
In fact, if the sign of the $\vec{d}$ vector is that of the direct type 
(as defined by Elhajal {\it et al.}~\cite{pyrochlores_elhajal_05}), 
the DM interaction also favors the all-in/all-out state and thus does not compete with the SIA. 
The competition arises for the opposite sign of the $\vec{d}$ vector, labelled as indirect type in \cite{pyrochlores_elhajal_05},  
where the DM couplings favor a continuous family of coplanar and non-coplanar GSs 
\cite{pyrochlores_elhajal_05}. 

{\it Conclusions.\,}
While trigonal distortions of the ligand cage are believed to open the door to topologically 
insulating states in $j\!=\!1/2$ 227 iridate pyrochlores \cite{227Ir_pesin_2010,227Ir_baek_kim_10,227Ir_savrasov_2011},
we here show that the lattice degrees of freedom can fundamentally modify the nature of 
the magnetic interactions and ground states in $5d$ 227 compounds with larger magnetic moments.
For this we employ {\it ab initio} multireference configuration-interaction methods,
the results of which are further mapped onto model Hamiltonians including both isotropic and anisotropic terms.
To extract the DM couplings, we have expanded the formalism proposed earlier
for $S\!=\!1/2$ cuprates \cite{DM_CuII_Maurice_2010,*DM_CuO_Pradipto_2012,*J_LiCu2O2_Pradipto_2012}.
For the experimentally determined crystal structure of Cd$_2$Os$_2$O$_7$ \cite{Os227_mandrus_2001}, 
we find an AF superexchange coupling $J\!=\!6.43$ meV and easy-axis anisotropy of the same magnitude, $D\!=\!-6.77$ meV, 
while the DM parameter is 3 to 4 times smaller. 
The dominant magnetic interactions, $J$ and $D$, give rise to the all-in/all-out AF order
\cite{227_moessner_98,pyrochlores_elhajal_05}. 
Most remarkably, the hexagonal Cd-ion coordination induces electrostatic fields that compete
with the compressive trigonal distortion of the O octahedra that is present in the system.
If the latter is removed the SIA turns positive, i.e., $D$ reverses sign and the system develops
easy-plane magnetic anisotropy. 
This obviously implies qualitatively different magnetic properties
\cite{hard_axis_Bramwell_94,*hard_axis_Champion_04}. 
Our findings thus indicate that there is in 227 pyrochlores a transition from unfrustrated
to highly frustrated magnetism controlled by the amount of trigonal distortion.
The latter can in principle be varied by changing either the $A$-site or $B$-site ions in
the $A_2B_2$O$_7$ compounds, replacing for instance Cd(Os) by Hg(Re) \cite{A2B2O7_rao_1983},
and is in addition tunable by applying pressure.
For a correct understanding of such magnetic interactions, ordering tendencies, and transitions
the subtle interplay between oxygen- and $A$-ion anisotropic electrostatic effects is essential.

{\it Acknowledgements.\,} We thank V.~M.~Katukuri, R.~Ganesh, H.~Stoll, and
P.~Fulde for insightful discussions.
This paper is dedicated to Prof.~Ria Broer to honour her devoted career in Theoretical and Computational Chemistry.
N.~A.~B. and L.~H. acknowledge financial support from the Erasmus Mundus Programme of the European Union
and the German Research Foundation (Deutsche Forschungsgemeinschaft, DFG), respectively.

\newpage

\setcounter{page}{1}
\setcounter{figure}{0}
\setcounter{table}{0}
\setcounter{equation}{0}
\renewcommand{\thefigure}{S\arabic{figure}}
\renewcommand{\thepage}{S\arabic{page}}
\renewcommand{\thetable}{S\Roman{table}}
\renewcommand{\theequation}{S\arabic{equation}}
\renewcommand{\bibnumfmt}[1]{[S#1]}
\renewcommand{\citenumfont}[1]{S#1}


\section{Supplementary material}

\section{Single-ion problem}

\begin{figure}[b!]
\includegraphics[width=0.8\columnwidth]{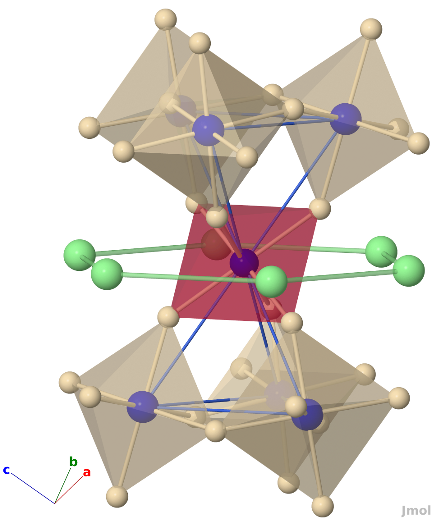}
\caption{
The cluster used for the ZFS calculations. Os, O, and Cd ions are shown in blue, beige, and green,
respectively.
}
\label{1scluster}
\end{figure}

To determine the Os $d$-level electronic structure we employed  
seven-octahedra clusters (see Fig.~\ref{1scluster}) embedded in arrays of point charges fitted to reproduce 
the crystal Madelung field in the cluster region.
Valence basis sets of quadruple-zeta quality \cite{SM_bookQC_2000} from the {\sc molpro} \cite{SM_molpro_brief} library 
were used for the central Os ion \cite{SM_ECP_Stoll_5d} and triple-zeta basis functions 
were applied for the ligands \cite{SM_GBas_molpro_2p} of the central octahedron
and the nearest-neighbor (NN) 5$d$ sites \cite{SM_ECP_Stoll_5d}.
For the central Os ion we also employed two polarization $f$ functions \cite{SM_ECP_Stoll_5d}.
For farther ligands in our clusters we applied minimal atomic-natural-orbital basis sets
\cite{SM_ANOs_pierloot_95}.
All occupied shells of the NN Cd$^{2+}$ ions were incorporated in the large-core pseudopotentials
and each of the Cd $5s$ orbitals was described by a single contracted Gaussian function \cite{SM_ECP_Dolg_4d}.
We accounted for all single and double excitations from the Os $5d$ and O $2p$ orbitals at the
central OsO$_6$ octahedron in the multireference configuration-interaction (MRCI) treatment.
To separate the O $2p$ valence orbitals into two different groups, i.e., at sites of the
central octahedron and at NN octahedra, we used the orbital localization module available in
{\sc molpro}.

We mapped our {\it ab initio} data onto a anisotropic single-site magnetic Hamiltonian.
The interaction matrix in the absence of magnetic fields is shown in Table~\ref{Hmod1} and
the effective Hamiltonian constructed on the basis of the MRCI+SOC data is given in Table~\ref{Heff1}.
The one-to-one correspondence between analogue terms in Tables~\ref{Hmod1} and~\ref{Heff1}
gives rise to a system of equations that can be easily solved to obtain
the single-ion anisotropy tensor in the original (crystallographic) coordinate frame. 
The later reads, in meV,
%
$$
\label{eq:Darb}
\bar{\bar{\vec{D}}}\!=\!\left(
\begin{array}{ccc}
 D_{xx} & D_{xy} & D_{xz} \\
 D_{xy} & D_{yy} & D_{yz} \\
 D_{xz} & D_{yz} & D_{zz}
\end{array} \right)\!=\!\left(
\begin{array}{ccc}
1.806 & -2.258 & -2.258 \\
-2.258 & 1.806 & -2.258 \\
-2.258 & -2.258 & 1.806
\end{array} \right)\!.
$$
The diagonalization of $\bar{\bar{\vec{D}}}$  yields the transformation matrix for rotating the initial coordinates to the 
magnetic framework and also the ZFS parameters, i.e., the axial $D$ and the rhombic $E$ components:
$$
\label{eq:Dmag}
\bar{\bar{\vec{D}}}_{\mathrm{mag}}\!=\!\left(
\begin{array}{ccc}
 D_{\mathrm{XX}} & 0 & 0 \\
 0 & D_{\mathrm{YY}} & 0 \\
 0 & 0 & D_{\mathrm{ZZ}}
\end{array} \right)\!=\!\left(
\begin{array}{ccc}
 4.064 & 0 & 0 \\
 0 & 4.064 & 0 \\
 0 & 0 & -2.710
\end{array} \right)\!,
$$
$$
D = D_{\mathrm{ZZ}}-\dfrac{1}{2}(D_{\mathrm{XX}}+D_{\mathrm{YY}})=-6.774 \text{ meV},
$$
$$
E = \dfrac{1}{2}(D_{\mathrm{XX}}-D_{\mathrm{YY}})=2\cdot 10^{-5} \text{ meV}\approx 0.
$$

\section{Two-site problem}

\begin{figure}[!t]
\includegraphics[width=1\columnwidth]{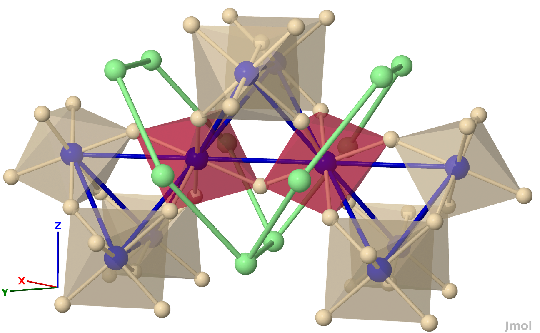}
\caption{
The cluster used for computing the intersite magnetic couplings. Os, O, and Cd sites are 
shown in blue, beige, and green, respectively.
The NN Os$^{5+}$ ions around the two central octahedra are modeled as Ta$^{5+}$ $d^0$ species.
}
\label{2scluster}
\end{figure}

Calculations for the inter-site magnetic couplings
were performed on embedded ten-octahedra clusters
with two active Os$^{5+}$ $5d^3$ sites, as shown in Fig.~\ref{2scluster}.
The basis functions used here were as described above.
The only exception was the O ligand bridging the two magnetically active Os sites, for which we employed
quintuple-zeta valence basis sets and four polarization $d$ functions \cite{SM_GBas_molpro_2p}.
The eight adjacent 5$d$ sites were modeled as Ta$^{5+}$ $d^0$ species.
Relative energies for the lowest singlet, triplet, quintet, and septet
are listed in Table~\ref{Jtable}.

The spin-orbit MRCI calculations and the subsequent analysis were performed in this
case within the manifold defined by a $t^3_{tg}$ $S\!=\!3/2$ electron configuration
at each active Os site. In other words, for this ${^4\!A_2}\!\otimes\!{^4\!A_2}$ manifold, we
do not include second-order couplings through higher lying crystal-field excited
states.
As a result, we describe only the diagonal interactions and the first-order
contribution to the Dzyaloshinskii-Moriya (DM) anisotropic term.
For 3$d$ oxides, it has been shown, nevertheless,
that the diagonal couplings are the most important, see Refs.~\cite{SM_DM_CuII_Maurice_2010,SM_DM_CuO_Pradipto_2012}.
For the same reason, the above treatment does not include the single-ion anisotropies,
which we have obtained from the single-site calculations described above.
Only $2p$ orbitals at the bridging O site were included in the MRCI treatment, as in
earlier CI calculations on $3d$ transition-metal compounds
\cite{SM_QC_J_fink94,*SM_J_ligand_calzado03,*SM_NOCI_J_hozoi03}.

\begin{table*}[t!]
\caption{Model interaction matrix for $\mathcal{\hat{H}}_{\mathrm{mod}} = \vec{S} \cdot \bar{\bar{\vec{D}}} \cdot \vec{S}$ and $S= \frac{3}{2}$.}
\label{Hmod1}
\begin{ruledtabular}
\begin{tabular}{rcccc}
$\mathcal{\hat{H}}_{\mathrm{mod}}$ & $\ket{\frac{3}{2},-\frac{3}{2}}$ & $\ket{\frac{3}{2},-\frac{1}{2}}$ & $\ket{\frac{3}{2},\frac{1}{2}}$ & $\ket{\frac{3}{2},\frac{3}{2}}$ \\

$\bra{\frac{3}{2},-\frac{3}{2}}$ & $ \frac{3}{4}(D_{xx}+D_{yy})+\frac{9}{4} D_{zz}$ & $-\sqrt{3} (D_{xz}+i D_{yz})$ & $\frac{\sqrt{3}}{2}  (D_{xx}-D_{yy}+2 i D_{xy})$
& $0$ \\
$\bra{\frac{3}{2},-\frac{1}{2}}$ & $ -\sqrt{3} (D_{xz}-i D_{yz})$ & $\frac{7}{4} (D_{xx}+D_{yy})+\frac{1}{4}D_{zz}$ & $0$ & $\frac{\sqrt{3}}{2}  (D_{xx}-D_{yy}+2 i D_{xy})$
\\
$\bra{\frac{3}{2},\frac{1}{2}}$ & $\frac{\sqrt{3}}{2}  (D_{xx}-D_{yy}-2 i D_{xy})$ & $0$ & $\frac{7}{4} (D_{xx}+D_{yy})+\frac{1}{4}D_{zz}$ & $\sqrt{3} (D_{xz}+i D_{yz})$ \\
$\bra{\frac{3}{2},\frac{3}{2}}$ &$0$ & $\frac{\sqrt{3}}{2}  (D_{xx}-D_{yy}-2 i D_{xy})$ & $\sqrt{3} (D_{xz}-i D_{yz})$ & $\frac{3}{4} (D_{xx}+D_{yy})+\frac{9}{4}D_{zz}$
\end{tabular}
\end{ruledtabular}
\end{table*}

We found that the {\it ab initio} results can be perfectly mapped onto a model Hamiltonian
with bilinear and biquadratic isotropic terms plus antisymmetric exchange, as presented
in Table~\ref{Hmod2}.
The effective Hamiltonian constructed using the MRCI+SOC data is shown
in Table~\ref{Heff2}.
Small non-diagonal elements on the order of 0.01 meV are neglected in
our analysis.
One-to-one correspondence between the elements in Tables~\ref{Hmod2} and
~\ref{Heff2} confirms that our choice  of $\mathcal{\hat{H^{\prime}}}_{\mathrm{mod}}$ is appropriate
and allows to extract all the numerical parameters of the model.
One should notice that
only the differences between the diagonal elements can be used in the analysis because
the trace of 
an effective Hamiltonian is in general arbitrary.

\begin{table*}[!b]
\caption{Single-site effective interaction matrix obtained by MRCI+SOC calculations on seven-octahedra cluster (meV).}
\label{Heff1}
\begin{ruledtabular}
\begin{tabular}{rcccc}
$\mathcal{\hat{H}}_{\mathrm{eff}}$ & $\ket{\frac{3}{2},-\frac{3}{2}}$ & $\ket{\frac{3}{2},-\frac{1}{2}}$ & $\ket{\frac{3}{2},\frac{1}{2}}$ & $\ket{\frac{3}{2},\frac{3}{2}}$ \\
$\bra{\frac{3}{2},-\frac{3}{2}}$ &$6.7738$            & $3.9110 + 3.9110 i$ & $-3.9106 i$          & $0.0000$ \\
$\bra{\frac{3}{2},-\frac{1}{2}}$ &$3.9110 - 3.9110 i$ & $6.7741$            & $0.0000$             & $-3.9106 i$ \\
$\bra{\frac{3}{2},\frac{1}{2}}$  &$3.9106 i$          & $0.0000$            & $6.7738$             & $-3.9110 - 3.9110 i$ \\
$\bra{\frac{3}{2},\frac{3}{2}}$  &$0.0000$            & $3.9106 i$          & $-3.9110 + 3.9110 i$ & $6.7738$
\end{tabular}
\end{ruledtabular}
\end{table*}

\begin{table}[!t]
\caption{
Relative energies (meV) of the singlet, triplet, quintet, and septet states for two adjacent Os $d^3$ sites.
In each case, the singlet is taken as reference.
}
\label{Jtable}
\begin{ruledtabular}
\begin{tabular}{cccc}
                                       &MCSCF   &MRCI      &MRCI+SOC \\
\hline
\\
$S_{\rm tot}\!=\!0$             &$0.00$  &$0.00$     &$0.00$    \\
$S_{\rm tot}\!=\!1$             &$2.51$  &$5.98$     &$6.04$    \\
                                       &         &           &$6.04$    \\
                                       &         &           &$6.44$    \\
$S_{\rm tot}\!=\!2$             &$7.75$  &$18.38$    &$18.66$    \\
                                       &         &           &$18.66$    \\
                                       &         &           &$19.01$    \\
                                       &         &           &$19.01$    \\
                                       &         &           &$19.12$    \\
$S_{\rm tot}\!=\!3$              &$16.24$ &$37.93$    &$38.51$    \\
                                       &         &           &$38.51$    \\
                                       &         &           &$38.81$    \\
                                       &         &           &$38.81$    \\
                                       &         &           &$38.99$    \\
                                       &         &           &$38.99$    \\
                                       &         &           &$39.05$    \\
\end{tabular}
\end{ruledtabular}
\end{table}

\begin{turnpage}
\squeezetable
\begin{table*}[p]
\caption{Model interaction matrix in coupled basis $\ket{S^{\mathrm{tot}},M_{S}^{\mathrm{tot}}}$ for $\mathcal{\hat{H^{\prime}}}_{\mathrm{mod}} = J \vec{S}_{1}  \cdot \vec{S}_{2} + K(\vec{S}_{1}  \cdot \vec{S}_{2})^2 + \vec{d}(\vec{S}_{1}  \times \vec{S}_{2} )$ and $S_{1,2}= \frac{3}{2}$}
\label{Hmod2}
\begin{ruledtabular}
\begin{tabular}{r|ccccccccc}
$\mathcal{\hat{H^{\prime}}}_{\mathrm{mod}}$ & $\ket{3,-3}$ & $\ket{3,-2}$ & $\ket{3,-1}$ & $\ket{3,0}$ & $\ket{3,1}$& $\ket{3,2}$ & $\ket{3,3}$  \\
\colrule
 $\bra{3,-3}$ & $\frac{9}{4}J+\frac{81}{16}K$ & 0 & 0 & 0 & 0 & 0 & 0 \\
 $\bra{3,-2}$ & 0 & $\frac{9}{4}J+\frac{81}{16}K$ & 0 & 0 & 0 & 0 & 0 \\
 $\bra{3,-1}$ & 0 & 0 & $\frac{9}{4}J+\frac{81}{16}K$ & 0 & 0 & 0 & 0 \\
 $\bra{3,0}$ & 0 & 0 & 0 & $\frac{9}{4}J+\frac{81}{16}K$ & 0 & 0 & 0 \\
 $\bra{3,1}$ & 0 & 0 & 0 & 0 & $\frac{9}{4}J+\frac{81}{16}K$ & 0 & 0 \\
 $\bra{3,2}$ & 0 & 0 & 0 & 0 & 0 & $\frac{9}{4}J+\frac{81}{16}K$ & 0 \\
 $\bra{3,3}$ & 0 & 0 & 0 & 0 & 0 & 0 & $\frac{9}{4}J+\frac{81}{16}K$ \\
 $\bra{2,-2}$ & $\frac{3}{2} \sqrt{\frac{3}{2}} (d_{y}+i d_{x})$ & $i\frac{3}{2}d_{z}$ &
 $\frac{3 }{2 \sqrt{10}}(d_{y}-i d_{x})$ & 0 & 0 & 0 & 0 \\
 $\bra{2,-1}$ & 0 & $\frac{3}{2} (d_{y}+i d_{x})$ & $i 3 \sqrt{\frac{2}{5}} d_{z}$ & 
$\frac{3}{2} \sqrt{\frac{3}{10}} (d_{y}-i d_{x})$ & 0 & 0 & 0 \\
 $\bra{2,0}$ & 0 & 0 & $\frac{3}{2} \sqrt{\frac{3}{5}} (d_{y}+i d_{x})$ & 
 $i \frac{9}{2 \sqrt{5}} d_{z}$ & $\frac{3}{2} \sqrt{\frac{3}{5}}(d_{y}-i d_{x})$ & 0 & 0 \\
 $\bra{2,1}$ & 0 & 0 & 0 & $\frac{3}{2} \sqrt{\frac{3}{10}} (d_{y}+i d_{x})$ & 
$i 3 \sqrt{\frac{2}{5}}  d_{z}$ & $\frac{3}{2}(dy_{y}-i d_{x} )$ & 0 \\
 $\bra{2,2}$ & 0 & 0 & 0 & 0 & $\frac{3 }{2 \sqrt{10}} (d_{y}+i d_{x})$& 
$i \frac{3 }{2} d_{z}$ & $\frac{3}{2} \sqrt{\frac{3}{2}}(d_{y}-i d_{x})$ \\
 $\bra{1,-1}$ & 0 & 0 & 0 & 0 & 0 & 0 & 0 \\
 $\bra{1,0}$ & 0 & 0 & 0 & 0 & 0 & 0 & 0 \\
 $\bra{1,1}$ & 0 & 0 & 0 & 0 & 0 & 0 & 0 \\
 $\bra{0,0}$ & 0 & 0 & 0 & 0 & 0 & 0 & 0 \\
\colrule \\
$\mathcal{\hat{H^{\prime}}}_{\mathrm{mod}}$ & $\ket{2,-2}$ & $\ket{2,-1}$ & $\ket{2,0}$ & $\ket{2,1}$ & $\ket{2,2}$ & $\ket{1,-1}$ & $\ket{1,0}$ & $\ket{1,1}$ & $\ket{0,0}$\\
\colrule
 $\bra{3,-3}$ & $\frac{3}{2} \sqrt{\frac{3}{2}} (d_{y}-i d_{x})$ & 0 & 0 & 0 & 0 & 0 & 
0 & 0 & 0 \\
 $\bra{3,-2}$ & $-\frac{3 }{2}i d_{z}$ & $\frac{3}{2} (d_{y}-i d_{x})$ & 0 & 0 & 0 & 0 & 
0 & 0 & 0 \\
 $\bra{3,-1}$ & $\frac{3}{2 \sqrt{10}} (d_{y}+i d_{x})$& $-i3 \sqrt{\frac{2}{5}} d_{z}$ &
 $\frac{3}{2} \sqrt{\frac{3}{5}} (d_{y}-i d_{x})$ & 0 & 0 & 0 & 0 & 0 & 0 \\
 $\bra{3,0}$ & 0 & $\frac{3}{2} \sqrt{\frac{3}{10}} (d_{y}+i d_{x})$ &
 $-i\frac{9 }{2 \sqrt{5}} d_{z}$& $\frac{3}{2} \sqrt{\frac{3}{10}}(d_{y}-i d_{x})$ & 0 &
 0 & 0 & 0 & 0 \\
 $\bra{3,1}$ & 0 & 0 & $\frac{3}{2} \sqrt{\frac{3}{5}} (d_{y}+i d_{x})$ &
 $-i 3 \sqrt{\frac{2}{5}} d_{z}$ & $\frac{3 }{2 \sqrt{10}} (d_{y}-i d_{x})$& 0 & 0 & 0 & 0 \\
 $\bra{3,2}$ & 0 & 0 & 0 & $\frac{3}{2} (d_{y}+i d_{x})$ & $-i\frac{3 }{2} d_{z}$& 0 & 0 & 0 & 0 \\
 $\bra{3,3}$ & 0 & 0 & 0 & 0 & $\frac{3}{2} \sqrt{\frac{3}{2}} (d_{y}+i d_{x})$ & 0 & 0 & 0 & 0 \\
 $\bra{2,-2}$ & $-\frac{3}{4}J+\frac{9}{16}K$ & 0 & 0 & 0 & 0 & $2 \sqrt{\frac{3}{5}} (d_{y}-i d_{x})$ & 0 & 0 & 0 \\
 $\bra{2,-1}$ & 0 & $-\frac{3}{4}J+\frac{9}{16}K$ & 0 & 0 & 0 & $-i2 \sqrt{\frac{3}{5}}  d_{z}$ &
 $\sqrt{\frac{6}{5}} (d_{y}-i d_{x})$ & 0 & 0 \\
 $\bra{2,0}$ & 0 & 0 & $-\frac{3}{4}J+\frac{9}{16}K$ & 0 & 0 & $\sqrt{\frac{2}{5}} (d_{y}+i d_{x})$ &
 $-i\frac{4 }{\sqrt{5}} d_{z}$& $\sqrt{\frac{2}{5}}(d_{y}-i d_{x})$ & 0 \\
 $\bra{2,1}$ & 0 & 0 & 0 & $-\frac{3}{4}J+\frac{9}{16}K$ & 0 & 0 & $\sqrt{\frac{6}{5}} (d_{y}+i d_{x})$ &
 $-i2 \sqrt{\frac{3}{5}} d_{z}$ & 0 \\
 $\bra{2,2}$ & 0 & 0 & 0 & 0 & $-\frac{3}{4}J+\frac{9}{16}K$ & 0 & 0 & $2 \sqrt{\frac{3}{5}} (d_{y}+i d_{x})$ & 0 \\
 $\bra{1,-1}$ & $2 \sqrt{\frac{3}{5}} (d_{y}+i d_{x})$ & $i2 \sqrt{\frac{3}{5}} d_{z}$ &
 $\sqrt{\frac{2}{5}} (d_{y}-i d_{x})$ & 0 & 0 & $-\frac{11}{4}J+\frac{121}{16}K$ & 0 & 0 & 
$\frac{1}{2} \sqrt{\frac{5}{2}} (d_{y}-i d_{x})$ \\
 $\bra{1,0}$ & 0 & $\sqrt{\frac{6}{5}} (d_{y}+i d_{x})$ & $i\frac{4}{\sqrt{5}}  d_{z}$ &
 $\sqrt{\frac{6}{5}} (d_{y}-i d_{x})$ & 0 & 0 & $-\frac{11}{4}J+\frac{121}{16}K$ & 0 & 
$-i\frac{1}{2} \sqrt{5}  d_{z}$ \\
 $\bra{1,1}$ & 0 & 0 & $\sqrt{\frac{2}{5}} (d_{y}+i d_{x})$ & $i2 \sqrt{\frac{3}{5}}  d_{z}$ &
 $2 \sqrt{\frac{3}{5}} (d_{y}-i d_{x})$
& 0 & 0 & $-\frac{11}{4}J+\frac{121}{16}K$ & $\frac{1}{2} \sqrt{\frac{5}{2}} (d_{y}+i d_{x})$ \\
 $\bra{0,0}$ & 0 & 0 & 0 & 0 & 0 & $\frac{1}{2} \sqrt{\frac{5}{2}} (d_{y}+i d_{x})$ &
 $i \frac{1}{2} \sqrt{5} d_{z}$ & $\frac{1}{2} \sqrt{\frac{5}{2}}(d_{y}-i d_{x})$ & $-\frac{15}{4}J+\frac{225}{16}K$
\end{tabular}
\end{ruledtabular}
\end{table*}
\end{turnpage}

\begin{turnpage}
\squeezetable
\begin{table*}[p]
\caption{Effective interaction matrix in coupled basis $\ket{S^{\mathrm{tot}},M_{S}^{\mathrm{tot}}}$, obtained from two-sites  MRCI+SOC calculations for Cd$_2$Os$_2$O$_7$ (meV)}
\label{Heff2}
\begin{ruledtabular}
\begin{tabular}{r|ccccccccc}
$\mathcal{\hat{H^{\prime}}}_{\mathrm{eff}}$ & $\ket{3,-3}$ & $\ket{3,-2}$ & $\ket{3,-1}$ & $\ket{3,0}$ & $\ket{3,1}$& $\ket{3,2}$ & $\ket{3,3}$  \\
\colrule
 $\bra{3,-3}$ &$38.525$ & $0.000$ & $0.016 i$ & $0.000$ & $-0.011$ & $0.000$ & $0.000$ \\
 $\bra{3,-2}$ & $0.000$ & $38.505$ & $0.000$ & $0.000$ & $0.000$ & $0.000$ & $0.000$  \\
 $\bra{3,-1}$ & $-0.016 i$ & $0.000$ & $38.511$ & $0.000$ & $0.010 i$ & $0.000$ & $-0.011$  \\
 $\bra{3,0}$ & $0.000$ & $0.000$ & $0.000$ & $38.507$ & $0.000$ & $0.000$ & $0.000$   \\
 $\bra{3,1}$ & $-0.011$ & $0.000$ & $-0.01 i$ & $0.000$ & $38.511$ & $0.000$ & $0.016 i$  \\
 $\bra{3,2}$ &  $0.000$ & $0.000$ & $0.000$ & $0.000$ & $0.000$ & $38.505$ & $0.000$\\
 $\bra{3,3}$ & $0.000$ & $0.000$ & $-0.011$ & $0.000$ & $-0.016 i$ & $0.000$ & $38.525$   \\
 $\bra{2,-2}$ & $-2.120+2.120 i$ & $0.000$ & $-0.550-0.550 i$ & $0.000$ & $0.000$ & $0.000$ & $0.000$   \\
 $\bra{2,-1}$ &  $0.000$ & $-1.727+1.727 i$ & $0.000$ & $-0.947-0.947 i$ & $0.000$ & $0.000$ & $0.000$  \\
 $\bra{2,0}$ & $0.000$ & $0.000$ & $-1.339+1.339 i$ & $0.000$ & $-1.339-1.339 i$ & $0.000$ & $0.000$  \\
 $\bra{2,1}$ &  $0.000$ & $0.000$ & $0.000$ & $-0.947+0.947 i$ & $0.000$ & $-1.727-1.727 i$ & $0.000$ \\
 $\bra{2,2}$ & $0.000$ & $0.000$ & $0.000$ & $0.000$ & $-0.550+0.550 i$ & $0.000$ & $-2.120-2.120 i$   \\
 $\bra{1,-1}$ & $0.000$ & $0.000$ & $0.000$ & $0.000$ & $0.000$ & $0.000$ & $0.000$   \\
 $\bra{1,0}$ &  $0.000$ & $0.000$ & $0.000$ & $0.000$ & $0.000$ & $0.000$ & $0.000$  \\
 $\bra{1,1}$ &  $0.000$ & $0.000$ & $0.000$ & $0.000$ & $0.000$ & $0.000$ & $0.000$  \\
 $\bra{0,0}$ &  $0.000$ & $0.000$ & $0.000$ & $0.000$ & $0.000$ & $0.000$ & $0.000$  \\
\colrule \\
$\mathcal{\hat{H^{\prime}}}_{\mathrm{eff}}$ & $\ket{2,-2}$ & $\ket{2,-1}$ & $\ket{2,0}$ & $\ket{2,1}$ & $\ket{2,2}$ & $\ket{1,-1}$ & $\ket{1,0}$ & $\ket{1,1}$ & $\ket{0,0}$\\
\colrule
 $\bra{3,-3}$ & $-2.120-2.120 i$ & $0.000$ & $0.000$ & $0.000$ & $0.000$ & $0.000$ & $0.000$ & $0.000$ & $0.000$ \\
 $\bra{3,-2}$ & $0.000$ & $-1.727-1.727 i$ & $0.000$ & $0.000$ & $0.000$ & $0.000$ & $0.000$ & $0.000$ & $0.000$ \\
 $\bra{3,-1}$ & $-0.550+0.550 i$ & $0.000$ & $-1.339-1.339 i$ & $0.000$ & $0.000$ & $0.000$ & $0.000$ & $0.000$ & $0.000$ \\
 $\bra{3,0}$ & $0.000$ & $-0.947+0.947 i$ & $0.000$ & $-0.947-0.947 i$ & $0.000$ & $0.000$ & $0.000$ & $0.000$ & $0.000$ \\
 $\bra{3,1}$ & $0.000$ & $0.000$ & $-1.339+1.339 i$ & $0.000$ & $-0.550-0.550 i$ & $0.000$ & $0.000$ & $0.000$ & $0.000$ \\
 $\bra{3,2}$ & $0.000$ & $0.000$ & $0.000$ & $-1.727+1.727 i$ & $0.000$ & $0.000$ & $0.000$ & $0.000$ & $0.000$ \\
 $\bra{3,3}$ & $0.000$ & $0.000$ & $0.000$ & $0.000$ & $-2.120+2.120 i$ & $0.000$ & $0.000$ & $0.000$ & $0.000$ \\
 $\bra{2,-2}$ & $18.954$ & $0.000$ & $-0.019 i$ & $0.000$ & $0.000$ & $1.827+1.827 i$ & $0.000$ & $0.000$ & $0.000$ \\
 $\bra{2,-1}$  & $0.000$ & $18.977$ & $0.000$ & $-0.023 i$ & $0.000$ & $0.000$ & $1.293+1.293 i$ & $0.000$ & $0.000$ \\
 $\bra{2,0}$ & $0.019 i$ & $0.000$ & $18.985$ & $0.000$ & $-0.019 i$ & $0.745-0.745 i$ & $0.000$ & $0.745+0.745 i$ & $0.000$ \\
 $\bra{2,1}$ & $0.000$ & $0.023 i$ & $0.000$ & $18.977$ & $0.000$ & $0.000$ & $1.293-1.293 i$ & $0.000$ & $0.000$ \\
 $\bra{2,2}$ & $0.000$ & $0.000$ & $0.019 i$ & $0.000$ & $18.954$ & $0.000$ & $0.000$ & $1.827-1.827 i$ & $0.000$ \\
 $\bra{1,-1}$ & $1.827-1.827 i$ & $0.000$ & $0.745+0.745 i$ & $0.000$ & $0.000$ & $6.528$ & $0.000$ & $0.000$ & $0.936+0.936 i$ \\
 $\bra{1,0}$  & $0.000$ & $1.293-1.293 i$ & $0.000$ & $1.293+1.293 i$ & $0.000$ & $0.000$ & $6.528$ & $0.000$ & $0.000$ \\
 $\bra{1,1}$  & $0.000$ & $0.000$ & $0.745-0.745 i$ & $0.000$ & $1.827+1.827 i$ & $0.000$ & $0.000$ & $6.528$ & $0.936-0.936 i$ \\
 $\bra{0,0}$ & $0.000$ & $0.000$ & $0.000$ & $0.000$ & $0.000$ & $0.936-0.936 i$ & $0.000$ & $0.936+0.936 i$ & $0.579$
\end{tabular}
\end{ruledtabular}
\end{table*}
\end{turnpage}


%
\end{document}